\documentclass[10pt,sigplan,twocolumn]{acmart}

\renewcommand\footnotetextcopyrightpermission[1]{}
\usepackage{amsmath,amssymb,amsfonts}
\usepackage{graphicx}
\usepackage{xcolor}
\usepackage{tabularray}
\UseTblrLibrary{booktabs}
\AtBeginDocument{%
  }

\acmConference[EuroSys '25]{Make sure to enter the correct
  conference title from your rights confirmation emai}{March 30--April 3,
  2025}{Rotterdam}
\acmISBN{978-1-4503-XXXX-X/18/06}

\begin{document}

\title{I Know What You Did Last Summer: Identifying VR User Activity Through VR Network Traffic}

\author{Sheikh Samit Muhaimin}
\email{smuhaimi@nd.edu}
\affiliation{%
  \institution{University of Notre Dame}
  \city{Notre Dame}
  \state{IN}
  \country{USA}
}

\author{Spyridon Mastorakis}
\email{mastorakis@nd.edu}
\affiliation{%
  \institution{University of Notre Dame}
  \city{Notre Dame}
  \state{IN}
  \country{USA}
}

\renewcommand{\shortauthors}{}

\begin{abstract}
  Virtual Reality (VR) technology has gained substantial traction and has the potential to transform a number of industries, including education, entertainment, and professional sectors. Nevertheless, concerns have arisen about the security and privacy implications of VR applications and the impact that they might have on users. In this paper, we investigate the following overarching research question: can VR applications and VR user activities in the context of such applications (e.g., manipulating virtual objects, walking, talking, flying) be identified based on the (potentially encrypted) network traffic that is generated by VR headsets during the operation of VR applications? To answer this question, we collect network traffic data from 25 VR applications running on the Meta Quest Pro headset and identify characteristics of the generated network traffic, which we subsequently use to train off-the-shelf Machine Learning (ML) models. Our results indicate that through the use of ML models, we can identify the VR applications being used with an accuracy of 92.4\% and the VR user activities performed with an accuracy of 91\%. Furthermore, our results demonstrate that an attacker does not need to collect large amounts of network traffic data for each VR application to carry out such an attack. Specifically, an attacker only needs to collect less than 10 minutes of network traffic data for each VR application in order to identify applications with an accuracy higher than 90\% and VR user activities with an accuracy higher than 88\%.
\end{abstract}

%




\maketitle

\section{Introduction}
Virtual Reality (VR) is rapidly permeating various aspects of daily life, including entertainment \cite{KIM2019346, JANG2019101239, 10.1007/978-3-319-28231-2_45}, education \cite{Checa2020-oe}, industry \cite{Berg2017-tl},and business \cite{9548495}. As VR applications become increasingly prevalent, the volume of network traffic that VR headsets generate also increases, thus a critical security and privacy concern emerges: can we identify what a VR user does in the context of a VR application (virtual environment) based on the network traffic that a VR headset generates?
As users interact with VR applications, they generate large amounts of data, including network traffic that contains information about packet transmissions, data exchanges, and communication patterns \cite{9539162}. Traditional traffic classification and activity recognition problems focus on network protocols and application data, while VR applications face challenges due to the diversity and amount of data generated by multiple unique sensors that include include haptic sensors, cameras, motion tracking etc. VR devices generate high-dimensional, time-sensitive data streams, making network transmission complicated. The heterogeneity and latency requirements make network management difficult, this calls into question whether the traffic analysis techniques currently in use are enough or if they should be modified or improved for circumstances peculiar to virtual reality. While the generation of network traffic generated by using VR applications is an unavoidable consequence of connecting VR users and allowing communication of VR headsets with VR infrastructure\footnote{There are VR applications that run mostly "locally" on a headset, however, such applications typically do not provide multi-user functionality and they are rather simplistic in nature. Applications that provide multi-user functionality typically generate large amounts of network traffic and these are the applications we focus on in this paper.}, it can also pose security and privacy risks if intercepted by malicious actors \cite{chauhan2024immersed, 10.1145/3597503.3639082, cheng2022we}. In this paper, we delve into the relationship between the network traffic generated by a VR headset during the operation of a VR application and the user behavior in a VR application. Our ultimate goal is to shed light on the feasibility of identifying VR user activities (e.g., throwing a virtual ball, painting, interacting with virtual objects) and application usage based on the generated network traffic.

Several key questions motivate our research, highlighting the security and privacy implications of user activity in virtual reality environments. These questions serve as the impetus for our research, prompting us to investigate the nuanced relationship between network traces and user behavior in VR applications. Specifically, in this paper, we aim to answer the following Research Questions (RQs): \textbf{RQ1.} Can we identify VR applications based on the generated network traffic?
\textbf{RQ2.} Can we identify user activity based on the generated network traffic?
\textbf{RQ3.} Can the same types of user activities be reliably identified across VR applications?
\textbf{RQ4.} Given a captured VR traffic trace generated by an unknown VR application, can we identify what user activities take place?
\textbf{RQ5.} “How much” VR traffic data do we need to capture in order to accurately identify VR applications and user activities?
\textbf{RQ6.} What are the performance and systems trade-offs of using machine learning for VR application and user activity identification?
By addressing these questions, our goal is to provide insights into the potential privacy risks associated with VR use.

To answer these research questions, in this paper, we collect network traffic traces generated by 25 Meta Quest Pro VR applications and we pursue the following approaches:  a. Identify correlations: We examine relationships between network traffic traces (such as TCP segments and UDP datagrams) and general user activities in VR apps. b. Detect variations: We identify variations in network trace data that correspond to specific user activities. c. Identification modeling: We investigate the feasibility of training Machine Learning (ML) models to identify VR user behavior based on VR network traffic.

In this paper, our main contributions\footnote{We will make our collected dataset and research artifacts publicly available for the research community to reuse with the camera ready version of our paper.} are the following:  

\begin {itemize}

\item We collect network traffic data generated by 25 VR applications running on the Meta Quest Pro headset under various settings and scenarios. We further identify features in the collected network traffic that can be used to answer the research questions posed above.  

\item We take advantage of ML models to identify VR applications and user activities based on the collected network traffic as well as investigate various trade-offs.

\end{itemize}

Our study demonstrates that we can accurately identify both the VR applications being used and the exact user activities within those applications based on the network traffic generated by VR headsets. More specifically, our results show that we can identify the VR applications used based on the generated network traffic with an accuracy of 92.4\%. Our results further show that we can identify specific user activities within the context of VR applications with an accuracy of 91\%. Additionally, our results demonstrated high precision and recall scores, indicating that we can effectively identify application usage and user activities across a variety of VR applications and scenarios.

\section{Problem and Motivation}

Interception and analysis of network traces present several opportunities for potential exploitation and intrusion in the context of VR \cite{9417659}. Malicious actors could use intercepted packet data to gain information about users' behavior, preferences, and activities in VR environments \cite{su2024remote}. 

\subsection{Threat Model}

We assume a threat model that allows malicious actors to deploy attacks against VR users in the wild. We describe the components of our threat model below.

\subsubsection{\textbf{Setting Up the Attack}}

We investigate a situation in which an attacker can utilize network traffic data obtained from VR applications to train ML models that identify user behavior and application usage. To intercept VR traffic traces, an attacker may place themselves on the same Wi-Fi network as the victim user, allowing them to capture network packets sent and received by VR headsets. Traffic of a VR headset in a captured traffic trace can be identified in a number of ways, for instance using metadata from network protocols, examining communication patterns distinctive to known VR applications, or analyzing packet payloads \cite{10.1145/3597503.3639082, trimananda2022ovrseen}. A specific example in the context of our study with the Meta Quest Pro headset is the destination Media Access Control (MAC) address of an Ethernet frame, which identifies the communication device. The destination field ``MetaPlat-6d:9d:eb (b4:17:a8:6d:9d:eb)'' refers to a Meta Platform device with the MAC address b4:17:a8:6d:9d:eb, which identifies that the device is a Meta VR headset. Another example in the context of our study is DNS. A DNS query to resolve a domain name ``graph.oculus.com'' is exclusive to Meta VR devices and associated services. Another such DNS query includes ``cdp.cloud.unity3d.com'', which identifies that applications use the Unity3D engine.

We further assume that: (i) VR generated traffic will be encrypted on an end-to-end basis as it is the case with commercially available headsets today \cite{10.1145/3597503.3639082}, therefore, an attacker will not have access to the actual packet payloads (i.e., application-layer data); and (ii) (at most) one VR application will be running on a headset at any time as it typically happens with  commercially available headsets today \cite{MetaForum2024}.

\begin{table}[htbp]
  \centering
  \caption{An overview of VR applications and user activities}
  \label{tab:activities_apps}
  \resizebox{\linewidth}{!}{
  \begin{tabular}{l|l}
    \toprule
    \textbf{VR Application} & \textbf{User Activity} \\
    \midrule
    Gym Class - BasketBall VR & No Activity, Paused, Ball Throwing, Talking, Walking \\ 
    Racket Fury & Talking, No Activity \\
    RecRoom & No Activity, Camera Change, Making A Fist, Walking \\
    Cards \& Tankards & Moving, Talking, Camera Change, Paused, No Activity \\
    Patchworld & No Activity, Paused, Button Pressing \\ Innerworld & Paused, Making A Fist, Walking, Talking, No Activity \\
    Remio & Flying, Talking, No Activity \\
    XRWorkout & Working-out, No Activity \\
    Multiverse & Walking, Talking, No Activity \\ ShapesXR & Creating Shapes, Interaction Mode, No Activity \\
    Big Ballers VR & Ball Spawning, No Activity \\
    Vegas Infinite & Moving, Talking, No Activity \\
    Gorilla Tag & Walking, Talking, No Activity \\
    Meta Horizon Workrooms & Drawing, Talking, No Activity \\
    Messenger & Talking, Texting, No Activity \\ Shootout & Shooting, Talking, Walking, No Activity \\
    Whatsapp & Talking, No Activity \\
    Noclip VR & Talking, Walking, No Activity \\
    Instagram & Talking, No Activity \\
    Monkey Doo & Throwing, Talking, No Activity \\
    First Person Tennis & Talking, No Activity \\ Wooorld & Moving, Talking, Navigating, No Activity \\
    Demeo & Talking, No Activity  \\
    Alcove & Position Change, Skin Change, Voice Command, , No Activity\\
    Meta Horizon Worlds- & Making A Fist, Walking, Paused, No Activity, \\
    (Earth Battling Gym) & Button Command \\

    \bottomrule
  \end{tabular}
  }
\end{table}

\subsubsection{\textbf{Relevant Data Collection}}

Once malicious actors gain access to the network packet data, they can use tools such as network protocol analyzers like Wireshark \cite{8319360} to record network packets from different VR apps to extract features for training. The selection of the 25 VR apps shown in TABLE \ref{tab:activities_apps}, was based on various requirements to ensure the relevance and diversity of the collected data. The chosen applications were those with a high volume of user activity and a wide range of interactions that could produce significant differences in network traffic. Apps that produced traffic only locally, without requiring network communication over the internet, were also disqualified. This exclusion was made to focus exclusively on applications where user actions directly affected network traffic, which is essential for analyzing traffic patterns and identifying user activities. The attacker begins gathering data about the communication patterns between VR headsets and external entities (e.g., servers, other VR users) which most probably will contain information such as source and destination IP addresses, packet sizes, packet timestamps, and the number of packets sent or received during a specific period of time.

\subsubsection{\textbf{Attacking Model(s)}}

By using the features of the network traffic data from collected traffic traces, off-the-shelf ML models, such as random forest, decision trees, and neural networks, can be trained to recognize user activity based on patterns of the observed network traffic. In other words, specific patterns of network traffic can be associated with activities like talking, interacting with virtual objects, navigating, or chatting with other users. For example, "Making a Fist" in the context of the "Meta Horizon Worlds (Earth Battling Gym)" application generates the maximum number of bits per second over five-second intervals in the generated traffic, so the models can easily distinguish this activity from other user activities. Data pre-processing, feature engineering, model selection, training, and evaluation are the steps that can be included in this method.

\subsubsection{\textbf{Re-Training (Fine Tuning) Models}}

Over time, an attacker may re-train (fine tune) the attacking model(s) in order to incorporate new VR applications and user activities, as well as incorporate new network traffic patterns. This aims to enhance the models' reliability and performance. Our threat model takes this fine tuning strategy into account.

\subsection{The Problem}

The type of attack we described above raises significant privacy concerns. By analyzing network packet data, attackers may gain insights which can expose users to a variety of cyber threats, including: 

\subsubsection{User Profiling} By analyzing network traffic, adversaries can create detailed profiles of users based on their VR activity, preferences, and interactions.

\subsubsection{Behavioral Tracking} Analyzing network packet data allows adversaries to track users' movements, actions, and interactions within VR applications, jeopardizing their privacy and anonymity.

\subsubsection{Social Engineering Attacks} Detailed profiles and behavioral tracking data can be used in social engineering attacks, enabling attackers to craft more convincing and targeted phishing attempts or other forms of social engineering.

\subsubsection{Data Breaches} Access to network traffic provides opportunities for attackers to exploit vulnerabilities in the VR applications, potentially leading to data breaches and the exposure of sensitive user information.

\subsubsection{Legal and Ethical Implications} Unauthorized access to network traffic data and the use of machine learning models for recognizing user activities raise serious legal and ethical concerns.

\subsection{Research Questions}

Below we motivate and elaborate on our research questions.
 
\noindent \textbf{RQ1: Can we accurately identify VR applications?} Our first aim is to explore whether the network traffic generated by VR headsets/applications exhibit unique patterns that can be used by attackers to identify the VR application that is running on a headset. In this research question, we investigate how well different ML models and methodologies distinguish different applications from their network traffic, given the complexity and diversity of network patterns across different VR applications. A list of the VR applications is shown in TABLE \ref{tab:activities_apps}.

\noindent \textbf{RQ2: Can we accurately identify user activity for a given VR application?} Unique network traffic patterns in VR applications can be generated by specific user activity. In this research question, we explore whether it is possible to precisely identify certain user activities (e.g., manipulating virtual objects, walking, flying, talking) in the context of a specific VR application based on the generated network traffic. This involves assessing the attacking ML models' accuracy and dependability. A list of all the user activities is demonstrated in TABLE \ref{tab:activities_apps}.

\begin{figure*}[!htbp]
\centering
\includegraphics[width=0.9\textwidth]{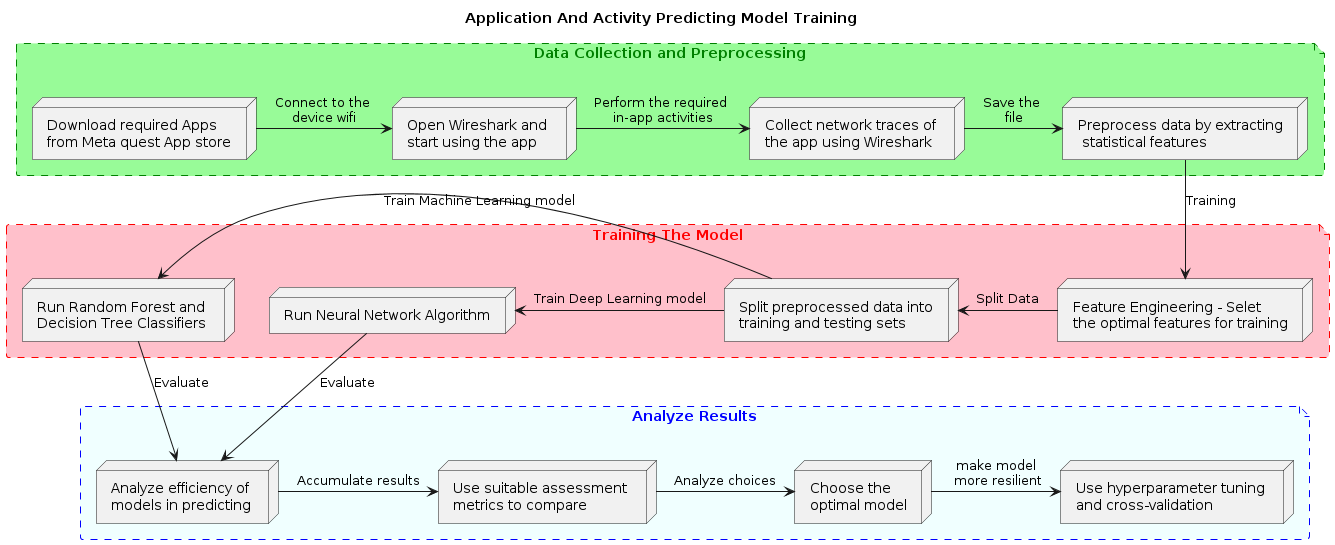}
\caption{Overview of methodology.}
\label{fig:method}
\end{figure*}

\noindent \textbf{RQ3: Can the same types of user activities be reliably identified across VR applications?} The consistency and generalizability of ML models in recognizing similar user behaviors across various VR applications are examined in this research question. In other words, we focus on user activities of the same type (e.g., walking, talking, moving around) that are present among several different VR applications and we aim to understand if ML models trained on a certain set of VR applications can consistently identify activities of the same type in a different set of VR applications.


\noindent \textbf{RQ4: Given a captured VR network traffic trace generated by an unknown VR application, can we identify what user activities take place?} In this research question, we investigate the possibility of deducing certain user behaviors from network traffic traces of an unknown VR application. To evaluate the ML model's generalizability and yield insightful information about user behavior, we employ ML models that have been trained on known VR applications to evaluate traffic traces from unknown VR applications and identify user behavior.

\noindent \textbf{RQ5: “How much” VR traffic data do we need to
capture in order to accurately identify VR applications
and user activities?} For precise identification, network traffic data of appropriate quantity and quality are essential. In this research question, our goal is to understand if there is some sort of ``ideal'' quantity of network traffic needed to accurately identify VR application and user behavior. Finding a balance that minimizes data collection efforts while increasing ML model performance involves examining the trade-offs between data volume, data granularity, and model accuracy.

\begin{figure}[htbp]
\centering
\includegraphics[width=0.5\linewidth]{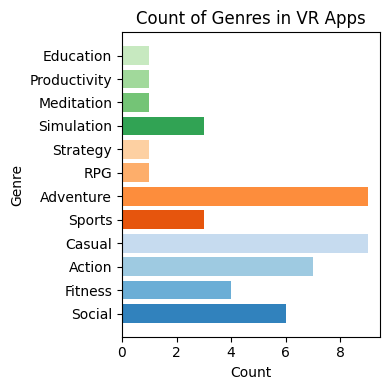}
\caption{Number of applications by genre}
\label{fig:count}
\end{figure}

\noindent \textbf{RQ6: What are the performance and systems trade-offs of using ML for VR application and user activity identification?} There are computational and system performance consequences associated with implementing ML models for the identification of VR applications and user behaviors. As a part of this research question, we want to quantify the impact of these models on system performance, including training and inference time, memory usage, and overall effectiveness. Additionally, we investigate the trade-offs among model accuracy, system resource consumption, and model complexity with the goal of maximizing identification efficiency within realistic computing constraints.

\section{Methodology and Results}
\subsection{Methodology}

We use a structured approach to construct and train ML models in order to identify VR user activity and application usage with the help of network packet data collected through Wireshark. On a high level, our methodology includes the phases of data preprocessing, feature engineering, model selection, training, and evaluation. We present an overview of our methodology in Fig. \ref{fig:method}.

\subsubsection{Data Collection} 

We use Wireshark to capture network packet data, which includes traffic generated by 25 Meta Quest Pro VR applications of different genres (Fig. \ref{fig:count}). We present an overview of the applications and the user activities for each application in Table \ref{tab:activities_apps}. 
We interacted with each application following the application's environment and flow. We extract the average packet length and delta time per second (i.e., the time between two consecutive network packets sent by a VR application) from all the packets that were captured using Wireshark. Every data collection session was intended to zero in on just one app at a time. This method was used to make sure that the traffic data gathered was unique to that app, preventing any cross-app interference. This technique makes it possible to analyze and model traffic patterns unique to each app with greater accuracy.
To guarantee that every task  was precisely identified and segregated, a manual approach was required. Although there is some noise in the labels, efforts were taken to reduce overlap by precisely timing the beginning and ending of every activity. The labeling noise is further reduced by the fact that all activities were performed by the same individual in a controlled environment during the data collection process. Subsequently, we combine them with data about the generated bits per second and packets per second with timestamps from Wireshark. Virtual reality applications offer new dimensions to user behavior and traffic patterns, but conventional traffic analysis fundamentals like packet sizes, inter-arrival durations, and transmission rate still provide valuable information. Preprocessing data is crucial for determining meaningful measuring criteria, such as maximum bits per second over a rolling window, and ensuring their effectiveness in virtual reality settings depends on customized techniques.

\subsubsection{Preprocessing Data}

The preprocessing step in identifying user behaviors in VR settings is more challenging than the model architecture itself. For example, one of the most important steps in capturing the brief periods of data transmission associated with particular user behaviors is figuring out the maximum bits per second during the last five seconds. This process required extensive trial and error and it ensures that the chosen time frame accurately captures the dynamic nature of virtual reality interactions, allowing for more accurate and reliable activity identification. The raw packet data was preprocessed to extract insights and create useful features for ML model training. We used the following strategies:

\begin{itemize}

\item \textbf{Extraction of Packet Length and Delta Time:} To obtain the temporal and spatial properties of network traffic, we calculated the average packet length and delta time per second.

\item \textbf{Standard Deviation Subtraction:} To standardize data and remove biases, we use the standard deviation subtraction technique, which facilitates the comparison and analysis of various network traffic patterns. First, the mean values of delta time, bits or packets per second, and packet length are obtained. To guarantee that the data is centered around zero, these mean values are then deducted from the matching feature values. This eliminates biases that could be brought about by different data scales or ranges. 

\item \textbf{Statistical Feature Generation:} To improve the richness of feature representation, we developed statistical features for several network traffic attributes over predetermined time intervals, including maximum, minimum, total, variance, and Cumulative Distribution Function (CDF) values. For each five-second interval, we calculated the maximum, minimum, total, variance, and CDF values of various features such as packet counts, delta time, and packet length.

\end{itemize}

\subsubsection{Feature Engineering} 

Based on empirical observations and experimentation, we established a collection of features that produced the best results for our ML models. These features included the following:
\begin{itemize}
    \item Number of packets per second.
    \item Standard deviation of time intervals between packets.
    \item Total sum of bits per second over five-second intervals.
    \item Time elapsed during the observation period in seconds.
    \item Total sum of all packets over five-second intervals.
    \item Minimum time interval between packets over five-second intervals.
    \item Distances to the nearest neighbors in the dataset.
    \item Minimum values observed among all packet counts over five-second intervals.
    \item Standard deviation of packet counts.
    \item Variance of bits per second counts.
    \item Variance of packet lengths.
\end{itemize}

\subsubsection{Model Selection}

Because of their adaptability and efficiency in processing a variety of datasets, we chose the classic random forest classifier and decision tree classifier as our ML models. In addition, we used a Multi-Layer Perceptron (MLP) neural network model, which can handle the multidimensional feature space generated by preprocessing VR traffic data. The MLP consists of an input layer, hidden layers using ReLU activation functions, and an output layer using a softmax function for activity classification. Generally, an 80-20 split was utilized, meaning that 80 percent of the data was used to train the models and 20 percent was set aside for testing. This split guarantees that a substantial percentage of the data is used for training the model, while also preserving a sufficient amount of untested data for assessment in order to avoid over-fitting and gauge the model's capacity for generalization.

\subsubsection{Evaluation}

The efficacy of the trained models in identifying VR user behavior and application usage was assessed through the use of suitable assessment metrics, including accuracy, precision, recall, and confusion matrices.  
To assess a model's effectiveness and scalability, measurements were also made of its training and inference time, CPU utilization, and memory consumption.

\subsubsection{Cross-validation and Hyperparameter Tuning}

We used methods like hyperparameter tuning and cross-validation to make sure our models were resilient. In addition to providing more accurate performance estimates and mitigating overfitting, hyperparameter tuning adjusted model parameters to improve identification accuracy.

\subsection{Results}

In this section, we attempt to address the research questions we initially put forward in order to effectively present the results of our study.

\begin{table}[htbp]
  \centering

    \caption{VR Application Identification Results}
    \label{tab:allappmet}
    \resizebox{0.9\linewidth}{!}{
      \begin{tabular}{lccc}
        \toprule
        \textbf{Metric} & \textbf{Random Forest} & \textbf{Decision Tree} & \textbf{Deep Learning} \\
        \midrule
        Accuracy & \textbf{92.4\%} & 87.1\% & 80.2\% \\
        Precision & \textbf{93\%} & 88\% & 83\% \\
        Recall & \textbf{92\%} & 87\% & 80\% \\
        \bottomrule
      \end{tabular}
    }
\end{table}

\subsubsection{\textbf{RQ1}} \textbf{Can we accurately identify VR applications?} 

\textbf{Insight 1.} \textit{Attackers can exploit unique patterns in network traffic generated by VR applications to train machine learning models that can accurately identify the specific VR application running on a headset.}

The random forest classifier with 300 trees achieved an application identification accuracy of 92.4\%, with a precision of 93\% and a recall of 92\%. These results indicate that the model can effectively identify the VR applications being used based on network traffic data as shown in Table \ref{tab:allappmet}. Precision and recall scores show that the model performs well in lowering false positives and false negatives, respectively. The model was able to identify patterns and correlations in the data by utilizing a large number of trees, which produced reliable and accurate identification of VR applications. 

\begin{table}[htbp]
  \centering
  \caption{The identification accuracy of different ML models for individual applications}
  \label{tab:classappaccu}
  \resizebox{0.9\columnwidth}{!}{
  \begin{tabular}{lccc}
    \toprule
    \textbf{Application class} & \textbf{Random Forest} & \textbf{Decision Tree} & \textbf{Deep Learning} \\
    \midrule
    Alcove & 78.9\% & 63.3\% & \textbf{100.0\%} \\
    Big Ballers VR & \textbf{100.0\%} & \textbf{100.0\%} & \textbf{100.0\%} \\
    Cards \& Tankards & 93.3\% & 83.0\% & \textbf{100.0\%} \\
    Demeo & \textbf{64.5\%} & 55.6\% & 38.0\% \\
    First Person Tennis & \textbf{100.0\%} & 95.2\% & 25.0\% \\
    Gorilla Tag & \textbf{100.0\%} & 90.9\% & \textbf{100.0\%} \\
    Gym Class BB VR & \textbf{100.0\%} & 96.6\% & \textbf{100.0\%} \\
    Innerworld & \textbf{100.0\%} & \textbf{100.0\%} & \textbf{100.0\%} \\
    Instagram & 92.4\% & 82.3\% & \textbf{100.0\%} \\
    Messenger & 93.0\% & 94.9\% & \textbf{96.0\%} \\
    MH Worlds EBG & \textbf{100.0\%} & \textbf{100.0\%} & \textbf{100.0\%} \\
    MH Workrooms & \textbf{100.0\%} & \textbf{100.0\%} & 0.0\% \\
    Monkey Doo & \textbf{94.4\%} & 88.9\% & 0.0\% \\
    Multiverse & \textbf{100.0\%} & 85.4\% & \textbf{100.0\%} \\
    Noclip VR & \textbf{98.3\%} & \textbf{98.3\%} & 0.0\% \\
    Patchworld & 97.7\% & 94.0\% & \textbf{100.0\%} \\
    Racket Fury TT VR & 91.3\% & 81.7\% & \textbf{94.0\%} \\
    RecRoom & 92.2\% & 91.6\% & \textbf{100.0\%} \\
    Remio & \textbf{96.4\%} & 92.8\% & 96.0\% \\
    ShapesXR & \textbf{94.2\%} & 71.7\% & 0.0\% \\
    Shootout & \textbf{100.0\%} & \textbf{100.0\%} & \textbf{100.0\%} \\
    Vegas Infinite & 81.2\% & 64.4\% & \textbf{100.0\%} \\
    Whatsapp & 86.5\% & 81.1\% & \textbf{100.0\%} \\
    Wooorld & 96.38\% & 95.39\% & \textbf{100.0\%} \\
    XRWorkout & 65.2\% & 63.0\% & \textbf{100.0\%} \\
    \bottomrule
  \end{tabular}
  }
\end{table}

The decision tree classifier demonstrated an accuracy of 87.1\% in application identification, a recall of 87\%, and a precision score of 88\%. In addition, the deep learning model achieved an accuracy of 80.2\%, a precision of 83\%, and a recall of 80\%. The results for the random forest classifier with 300 trees were better than the decision tree classifier and the deep learning model, since random forest averages the output of a number of trees. This minimizes overfitting and results in identification decisions that are more reliable and accurate.

\begin{table}[htbp]
  \centering
  \caption{Identification accuracy for individual VR user activities per VR application. Results for Random Forest (RF), Decision Tree (DT), and Deep Learning (DL) are presented(Part-1).}
  \label{tab:pairactaccu_combined1}
  \resizebox{0.9\columnwidth}{!}{
  \begin{tabular}{llccc}
    \toprule
    \textbf{Activity} & \textbf{App} & \textbf{RF} & \textbf{DT} & \textbf{DL} \\
    \midrule
    Button Pressing & Patchworld & \textbf{100\%} & \textbf{100\%} & 90\% \\
    No Activity & Noclip VR & \textbf{86.96\%} & 65.22\% & 0.00\% \\ 
    No Activity & Patchworld & 94.37\% & 87.32\% & \textbf{95.77\%} \\ 
    Talking & Noclip VR & \textbf{100\%} & 96.08\% & 0.00\% \\ 
    Paused & Patchworld & \textbf{100\%} & 95.45\% & \textbf{100\%} \\ Walking & Noclip VR & \textbf{82.76\%} & 39.66\% & 0.00\% \\ 
    No Activity & MH Workrooms & \textbf{95.77\%} & 92.96\% & 0.00\% \\
    No Activity & Monkey Doo & \textbf{85.88\%} & 78.82\% & 0.00\% \\ 
    Talking & MH Workrooms & \textbf{98.18\%} & \textbf{98.18\%} & 0.00\% \\
    Throwing & Monkey Doo & 73.91\% & \textbf{78.26\%} & 0.00\% \\ 
    Drawing & MH Workrooms & \textbf{100\%} & 93.88\% & 0.00\% \\ Talking & Monkey Doo & \textbf{97.96\%} & 75.51\% & 0.00\% \\ 
    Talking & Remio & 91.84\% & 83.67\% & \textbf{100\%} \\ 
    No Activity & MH Worlds EBG & 93.10\% & 89.66\% & \textbf{94.83\%} \\ 
    Flying & Remio & \textbf{80.43\%} & 56.52\% & 78.26\% \\
    Paused & MH Worlds EBG & \textbf{96.15\%} & \textbf{96.15\%} & 84.62\% \\ 
    No Activity & Remio & \textbf{84.09\%} & 65.91\% & 68.26\% \\ Making A Fist & MH Worlds EBG & \textbf{100\%} & 61.90\% & 90.48\% \\ 
    Position Change & Alcove & \textbf{97.67\%} & 93.02\% & 79.07\% \\
    No Activity & RecRoom & \textbf{83.78\%} & 67.57\% & 79.73\% \\ 
    Skin Change & Alcove & 80.00\% & \textbf{82.00\%} & \textbf{82.00\%} \\
    Camera Change & RecRoom & 55.81\% & \textbf{60.47\%} & 44.19\% \\ 
    Voice Command & Alcove & 76.92\% & \textbf{96.15\%} & 76.92\% \\
    Making A Fist & RecRoom & 41.67\% & 27.78\% & \textbf{69.44\%} \\ 
    Button Command & Alcove & \textbf{79.31\%} & 32.76\% & 29.31\% \\
    Walking & RecRoom & \textbf{61.54\%} & 42.31\% & 34.62\% \\ 
    No Activity & Alcove & 49.68\% & 35.48\% & \textbf{74.19\%} \\ No Activity & Gym Class BB VR & 91.72\% & 84.83\% & \textbf{92.07\%} \\ 
    Creating Shapes & ShapesXR & \textbf{92.00\%} & 40.00\% & 0.00\% \\ 
    Paused & Gym Class BB VR & 79.59\% & 20.41\% & \textbf{100\%} \\ 
    Interaction Mode & ShapesXR & \textbf{97.96\%} & 71.43\% & 0.00\% \\
    Ball Throwing & Gym Class BB VR & \textbf{100\%} & \textbf{100\%} & \textbf{100\%} \\ 
    No Activity & ShapesXR & \textbf{79.73\%} & 64.86\% & 0.00\% \\ Talking & Gym Class BB VR & 71.60\% & \textbf{75.31\%} & 54.32\% \\ 
    No Activity & Messenger & 93.64\% & 83.64\% & \textbf{95.45\%} \\
    Walking & Gym Class BB VR & \textbf{98.68\%} & \textbf{98.68\%} & 22.37\% \\ 
    Talking & Messenger & \textbf{93.75\%} & 85.42\% & \textbf{93.75\%} \\
    No Activity & Cards \& Tankards & \textbf{87.61\%} & 75.22\% & 78.76\% \\ 
    No Activity & Whatsapp & \textbf{78.05\%} & 65.85\% & 63.41\% \\
    Moving & Cards \& Tankards & \textbf{85.71\%} & 17.14\% & 14.29\% \\ 
    Talking & Whatsapp & \textbf{96.97\%} & \textbf{96.97\%} & 12.12\% \\
    Talking & Cards \& Tankards & 70.59\% & 39.22\% & \textbf{100\%} \\
    Talking & Vegas Infinite & 77.61\% & 67.16\% & \textbf{100\%} \\ Paused & Cards \& Tankards & 96.23\% & 81.13\% & \textbf{100\%} \\ 
    Moving & Vegas Infinite & 46.67\% & 16.67\% & \textbf{83.33\%} \\ 
    No Activity & Shootout & \textbf{98.84\%} & 80.23\% & 83.72\% \\ 
    No Activity & Vegas Infinite & 79.58\% & 61.27\% & \textbf{84.51\%} \\
    Talking & Shootout & \textbf{93.44\%} & 85.25\% & \textbf{93.44\%} \\ 
    No Activity & XRWorkout & \textbf{79.67\%} & 73.98\% & 80.49\% \\
    Shooting & Shootout & 84.62\% & \textbf{88.46\%} & 51.92\% \\ 
    Working-out & XRWorkout & \textbf{59.26\%} & 9.26\% & 42.59\% \\
    Walking & Shootout & 68.00\% & 68.00\% & \textbf{72.00\%} \\ 
    Talking & XRWorkout & 28.30\% & 37.74\% & \textbf{56.60\%} \\ Moving & Woorld & 46.00\% & \textbf{78.00\%} & 58.00\% \\ 
    No Activity & Racket Fury TT VR & \textbf{92.65\%} & \textbf{92.65\%} & 88.24\% \\
    Talking & Woorld & \textbf{100\%} & 92.16\% & 90.20\% \\
    Talking & Racket Fury TT VR & 86.11\% & 55.56\% & \textbf{100\%} \\
    Navigating & Woorld & 82.46\% & 73.68\% & \textbf{85.09\%} \\
     No Activity & Instagram & 84.62\% & 69.23\% & \textbf{87.18\%} \\
    No Activity & Woorld & \textbf{85.39\%} & 78.65\% & 83.15\% \\ 
    Talking & Instagram & 92.50\% & 80.00\% & \textbf{100\%} \\
    No Activity & Gorilla Tag & 96.08\% & 96.08\% & \textbf{94.12\%} \\ 
    No Activity & Innerworld & \textbf{100\%} & \textbf{100\%} & 69.47\% \\
    Walking & Gorilla Tag & \textbf{100\%} & \textbf{100\%} & \textbf{100\%} \\ 
    \bottomrule
  \end{tabular}
  }
\end{table}

\begin{table}[htbp]
  \centering
  \caption{Identification accuracy for individual VR user activities per VR application. Results for Random Forest (RF), Decision Tree (DT), and Deep Learning (DL) are presented (Part-2).}
  \label{tab:pairactaccu_combined2}
  \resizebox{0.7\columnwidth}{!}{
  \begin{tabular}{llccc}
    \toprule
    \textbf{Activity} & \textbf{App} & \textbf{RF} & \textbf{DT} & \textbf{DL} \\
    \midrule

    Talking & Innerworld & \textbf{100\%} & \textbf{100\%} & 76.92\% \\
    Talking & Multiverse & \textbf{100\%} & 94.94\% & 98.73\% \\ 
    Walking & Innerworld & \textbf{100\%} & \textbf{100\%} & 86.96\% \\
    No Activity & Multiverse & \textbf{88.64\%} & 70.45\% & 86.36\% \\ 
    Paused & Innerworld & \textbf{100\%} & \textbf{100\%} & 78.43\% \\

\bottomrule
  \end{tabular}
  }
\end{table} 

\begin{figure*}[htbp]
\centering
\includegraphics[width=0.7\textwidth]{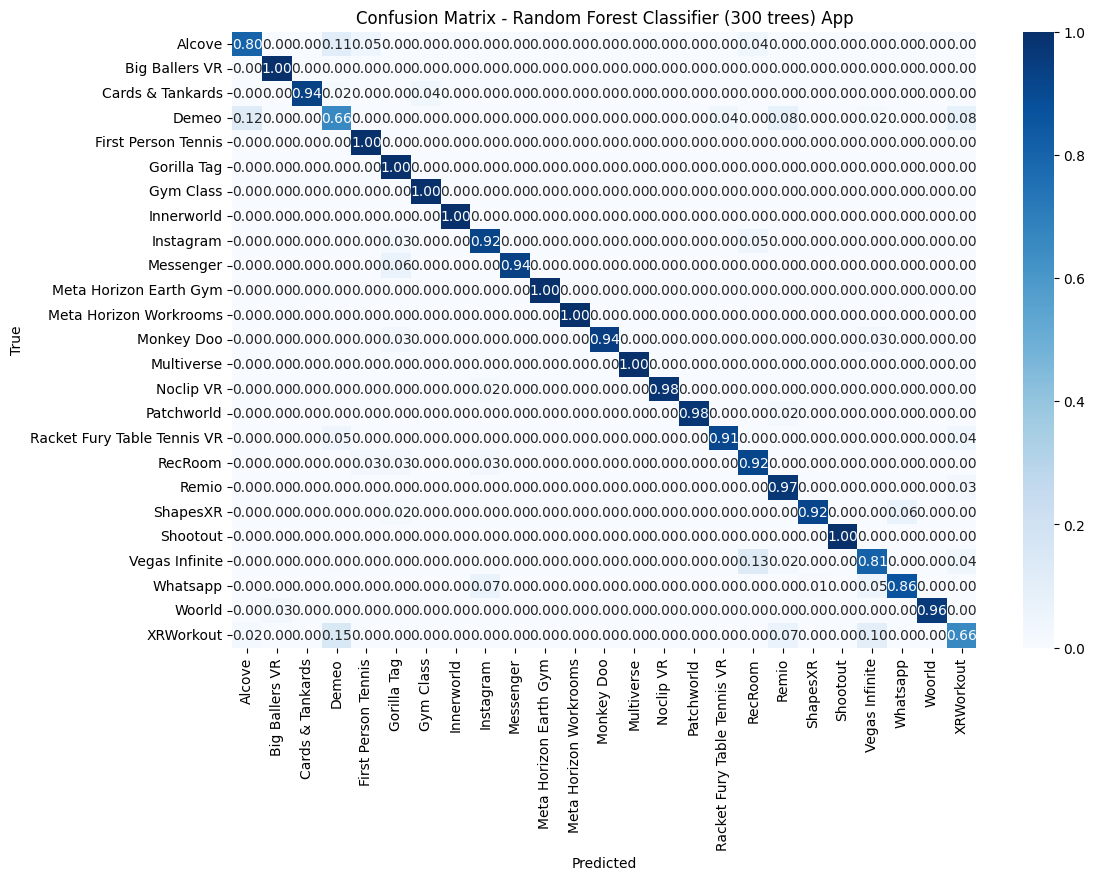}
\caption{Confusion matrix for the identification of VR applications through a random forest classifier.}
\label{fig:conf_appreal}
\end{figure*}

In Table~\ref{tab:classappaccu}, we present the accuracy for individual VR applications and for each different type of ML classifier. Random forest achieves the highest accuracy for the majority of VR applications. In addition, the random forest classifier achieves a perfect accuracy for several applications. 
Despite having a satisfactory overall performance, the decision tree classifier frequently performs worse than the other classifiers. For example, the accuracy of the decision tree classifier is worse in applications, such as Demeo and First Person Tennis, due to its tendency to overfit when noise is present in the training set. However, because deep learning can represent intricate, non-linear connections, it can also often achieve 100\%, which is useful for identifying a variety of patterns. However, deep learning absolutely fails (0\% accuracy) in other applications, such Monkey Doo and MH Workrooms, suggesting potential problems with data representation or model generalization. In scenarios with less complex data and highly specific patterns that are easier to capture with simpler models, the deep learning model underperforms.

Considering that the random forest classifier achieves the highest performance, we present its confusion matrix in Figure~\ref{fig:conf_appreal}. The confusion matrix provides detailed information about the random forest classifier's performance for each VR application, including instances of misclassifications and the classes (applications) they are confused with. Although there were a few instances of confusion in the model's classification process, such as in the cases of Alcove and XRWorkout VR applications, the random forest classifier was able to correctly identify the vast majority of VR applications in a consistent manner.

\begin{figure*}[htbp]
\centering
\includegraphics[width=0.65\linewidth]{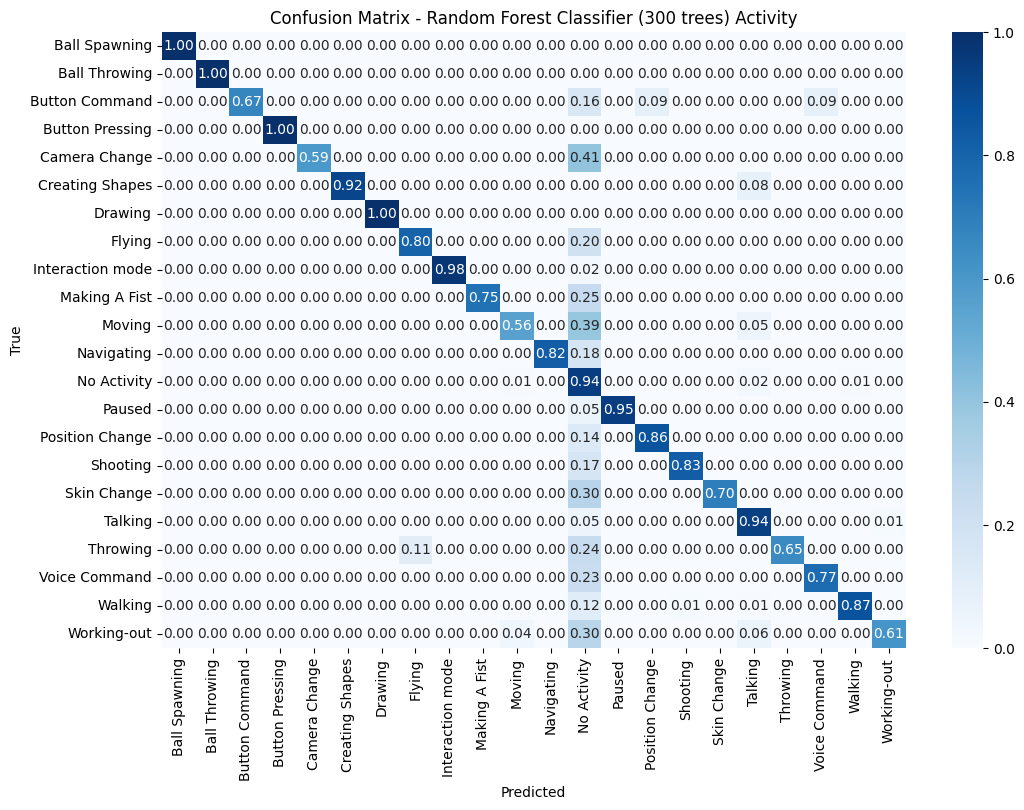}
\caption{Confusion matrix for various user activities regardless of applications in the case of the random forest classifier.}
\label{fig:conf_acc}
\end{figure*}

\subsubsection{\textbf{RQ2}} \textbf{Can we accurately identify user activity for a given VR application?} 

\textbf{Insight 2.} \textit{We can identify particular user actions taking place within a VR application with high accuracy by utilizing the network traffic patterns that the application generates.}

The random forest classifier with 300 trees achieved an activity identification accuracy of 91.2\% with a precision of 91\% and a recall of 91\%, demonstrating its ability to accurately identify user activities within VR applications as presented in Table~\ref{tab:allactmet}. At the same time, the decision tree classifier achieved an identification accuracy of 82\% in VR user activity classification. The user activity identification recall rate was 82\% and the precision score was 83\%. The user activity identification accuracy for deep learning was 77.5\% with a precision of 75\% and a recall of 78\%. Overall, the random forest classifier achieved the best identification results for user activities by averaging the output of multiple trees in order to reduce overfitting.

\begin{table}[htbp]

    \centering
    \caption{VR User Activity Identification Results}
     \label{tab:allactmet}
    \resizebox{0.8\linewidth}{!}{
      \begin{tabular}{lccc}
        \toprule
        \textbf{Metric} & \textbf{Random Forest} & \textbf{Decision Tree} & \textbf{Deep Learning} \\
        \midrule
        Accuracy & \textbf{91.2\%} & 82\% & 77.5\% \\
        Precision & \textbf{91\%} & 83\% & 75\% \\
        Recall & \textbf{91\%} & 82\% & 78\% \\
        \bottomrule
      \end{tabular}
    }
  
\end{table}

The accuracy of random forest, decision tree, and deep learning models for each user action in the context of each different VR application is presented in Tables~\ref{tab:pairactaccu_combined1} and \ref{tab:pairactaccu_combined2}. The random forest classifier consistently performs better than the two other models across a range of actions and applications, demonstrating its ability to generalize the performed identification tasks to a variety of user activities and applications. Specifically, there are several VR user activities and applications, where random forest can achieve a perfect identification accuracy, which is more than any other model we tried.

Even with its decent performance, the decision tree classifier frequently trails behind random forest. As an example, the decision tree classifier achieves an accuracy of 56.52\% during the ``flying'' activity in the context of the ``Remio'' VR application, a considerable decrease from random forest's 80.43\%. This is explained by the tendency of the decision tree classifier to overfit, which causes it to capture noise instead of underlying patterns.

\begin{table}[htbp]
  \centering
  \caption{Overall activity accuracy comparison before and after removing "no activity".}
  \label{tab:noactall_acc_comparison}
  \resizebox{1\columnwidth}{!}{
  \begin{tabular}{lcc}
    \toprule
    \textbf{Model} & \textbf{Original accuracy} & \textbf{Accuracy after removing "No Activity"} \\
    
    \midrule
    Random Forest & 91.2\% & \textbf{96.16\%} \\
    Decision Tree & 82\% & \textbf{91\%} \\
    Deep Learning & 77.5\% & \textbf{79.4\%} \\
    \bottomrule
  \end{tabular}
  }
\end{table}

Deep learning, in general, performed worse than random forest and decision tree mainly due to the fact that it requires larger volumes of data than the machine learning models in order to learn effective complex patterns and characteristics of the training dataset. The outcomes highlight how crucial context is to the model's functionality. 
All models reach 100\% accuracy in dynamic activities like "Walking" in "Gorilla Tag," demonstrating their capacity to handle movement data. On the other hand, in "Alcove," during the "Voice Command" task, decision tree attains the maximum accuracy of 96.15\%, suggesting that it is appropriate for voice-based interactions where rule-based decisions are easier to understand.

For every activity, the confusion matrix in Figure~\ref{fig:conf_acc} offers comprehensive details about the random forest classifier's performance, including instances of misclassification and the classes they are confused with. The confusion matrix shows that the model performed well overall across a variety of user activities, resulting in some instances of misclassification for certain activities. 

\begin{table*}[htbp]
\centering
\caption{Activity accuracy across applications for Random Forest (RF), Decision Tree (DT), and Deep Learning (DL) before and after removing "no activity" (improved after removing "no activity").}
\label{tab:combined_noactacc_comparison}
\resizebox{0.6\textwidth}{!}{
\begin{tabular}{llllllll}
\toprule
\textbf{Activity Class} & \textbf{RF (Initial)} & \textbf{RF (Improved)} & \textbf{DT (Initial)} & \textbf{DT (Improved)} & \textbf{DL (Initial)} & \textbf{DL (Improved)} \\
\midrule
Ball Spawning & 97.70\% & \textbf{100\%} & 88.64\% & \textbf{100\%} & 98\% & \textbf{100\%} \\ 
Ball Throwing & \textbf{100\%} & \textbf{100\%} & \textbf{100\%} & \textbf{100\%} & \textbf{100\%} & \textbf{100\%} \\ 
Button Command & 62.10\% & \textbf{91.38\%} & \textbf{51.7\%} & \textbf{51.7\%} & 24\% & \textbf{71\%} \\ 
Button Pressing & \textbf{100\%} & \textbf{100\%} & 62.5\% & \textbf{100\%} & \textbf{100\%} & \textbf{100\%} \\ 
Camera Change & 55.40\% & \textbf{100\%} & 70.27\% & \textbf{89.19\%} & 22\% & \textbf{64\%} \\ 
Creating Shapes & \textbf{92.00\%} & 90.00\% & 90\% & \textbf{92\%} & 0\% & \textbf{48\%} \\ 
Drawing & \textbf{100\%} & \textbf{100\%} & 93.88\% & \textbf{100\%} & \textbf{0\%} & \textbf{0\%} \\ 
Flying & 84.80\% & \textbf{89.13\%} & 28.26\% & \textbf{89.13\%} & 67\% & \textbf{100\%} \\ 
Interaction mode & 98\% & \textbf{100\%} & 87.75\% & \textbf{100\%} & 0\% & \textbf{100\%} \\ 
Making A Fist & 74.70\% & \textbf{92.77\%} & 68.67\% & \textbf{98.8\%} & 76\% & \textbf{92\%} \\ 
Moving & 70.80\% & \textbf{86.15\%} & 60\% & \textbf{78.26\%} & 33\% & \textbf{66\%} \\ 
Navigating & 83.33\% & \textbf{86.84\%} & 76.32\% & \textbf{86.84\%} & 72\% & \textbf{81\%} \\ 
Paused & 95.50\% &\textbf{99.50\%} & 76.62\% &\textbf{97.01\%} & 93\% &\textbf{98\%} \\ 
Position Change & 86\% & \textbf{100\%} & 53.49\% & \textbf{88.37\%} & 84\% & \textbf{93\%} \\ 
Shooting & 84.60\% & \textbf{100\%} & 78.85\% & \textbf{100\%} & 48\% & \textbf{67\%} \\ 
Skin Change & 80\% & \textbf{90\%} & 80\% & \textbf{70.00\%} & 72\% & \textbf{80\%} \\ 
Talking & 93.40\% & \textbf{98.73\%} & 83.97\% & \textbf{92.58\%} & 82\% & \textbf{85\%} \\ 
Throwing & 76.10\% & \textbf{89.13\%} & 78.26\% & \textbf{86.96\%} & \textbf{0\%} & \textbf{0\%} \\ 
Voice Command & 76.90\% & \textbf{100\%} & 96.15\% & \textbf{100\%} & 77\% & \textbf{81\%} \\ 
Walking & 87.80\% & \textbf{95.72\%} & 78.62\% & \textbf{93.75\%} & 64\% & \textbf{68\%} \\ 
Working-out & 66.70\% & \textbf{79.63\%} & 18.52\% & \textbf{68.51\%} & 50\% & \textbf{93\%} \\
\bottomrule
\end{tabular}
}

\end{table*}

\subsubsection{\textbf{RQ3}} \textbf{Can the same types of user activities be reliably identified across VR applications?} 

\textbf{Insight 3.} \textit{The consistent identification of similar activity types across various applications by the models can be attributed to the uniform patterns in network traffic data, likely consequence of the frequent use of similar game engines in VR development.}

Our results in Table \ref{tab:combined_noactacc_comparison} show that we can identify the same types of user activities (e.g., ``walking'', ``making a fist'', ``talking'') regardless of the application being used. Specifically, as we show in Table~\ref{tab:noactall_acc_comparison}, the overall accuracy for random forest is about 91\%, the overall accuracy for the decision tree classifier is about 82\%, and the overall accuracy for deep learning is about 77.5\%. In addition, Figure~\ref{fig:conf_acc} shows the confusion matrix of our random forest classifier when we attempt to identify user activities regardless of applications. Our results show that several activities can be identified with an accuracy greater than 90\%. However, several activities are also confused with cases of ``no activity'', which are instances of the user being essentially idle. This is due to the fact that such idle instances resemble the structure of noise, thus essentially adding noisy patterns into the overall dataset.

Considering that such instances of VR users being idle do not provide context on the actual activities performed users, we further evaluated the impact of omitting such idle cases from our dataset. Our results in Tables~\ref{tab:combined_noactacc_comparison} and~\ref{tab:noactall_acc_comparison} demonstrate that omitting the noise patterns introduced into the overall dataset by instances of no user activity improve the overall accuracy of all classifiers as well as the accuracy of all classifiers for individual activities. Specifically, the accuracy for individual activities was increased by up to 44\%, while the overall accuracy of classifiers was increased by up to 9\%. 

Additionally, VR applications use common game engines like FMOD which standardizes audio processing by using similar algorithms for sound effects, mixing, and spatialization across applications. This leads to consistent audio data transmission patterns, which can be reflected in the network traffic, making it easier to identify similar user behaviors across different VR apps.

\begin{figure}[htbp]
\centering
\includegraphics[width=\linewidth]{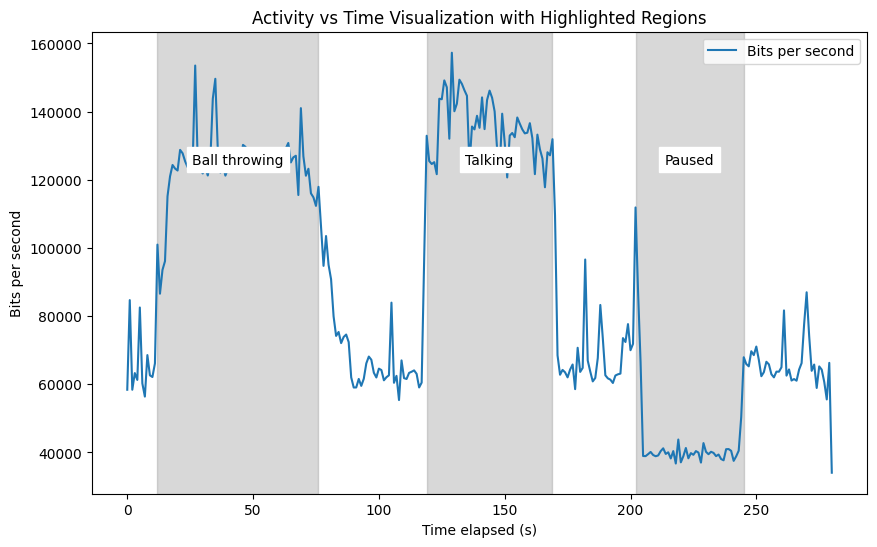}
\caption{An example of identification of VR user activities in a traffic trace containing unknown traffic data.}
\label{fig:activityvstime}
\end{figure}

\begin{figure*}[htbp]
\centering
\includegraphics[width=0.7\textwidth]{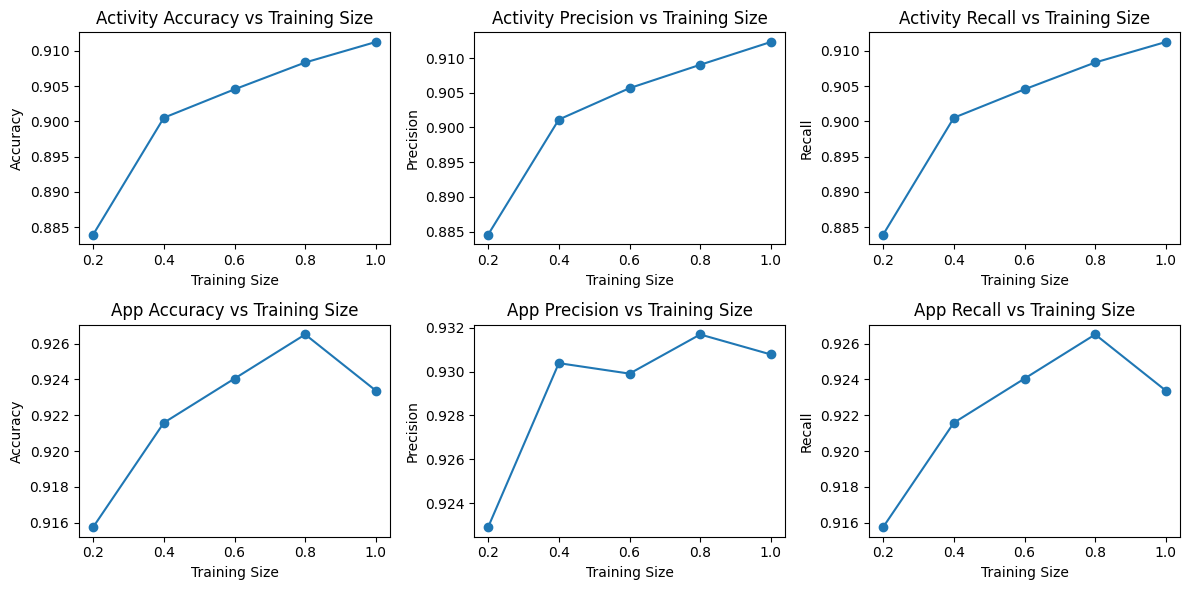}
\caption{Accuracy, precision, and recall results for VR application and activity identification for different amounts of training data.}
\label{fig:actacc}
\end{figure*}

\subsubsection{\textbf{RQ4}} \textbf{Given a captured VR traffic trace generated by an unknown VR application, can we identify what user activities take place?}

\textbf{Insight 4.} \textit{The model's capacity for generalization accounts for its accuracy in identifying user behaviors from network traffic traces associated with an unknown VR application.}

In Figure~\ref{fig:activityvstime}, we present an example of an unknown VR application traffic trace, where our models identify VR user activities as they take place over time. For this specific traffic trace, our models have identified three activities (throwing a ball, talking to other players, and pausing the application). As we previously discussed, our results in Table~\ref{tab:allactmet} indicate that user activity can be determined within an application using the three models we have selected. Specifically, activities can be detected with up to 91\% of accuracy, precision, and recall by a random forest classifier. These results still hold even when the classifiers are presented with unknown traffic data.Tables~\ref{tab:pairactaccu_combined1}, \ref{tab:pairactaccu_combined2}, and~\ref{tab:combined_noactacc_comparison} further demonstrate that most applications can be accurately identified on an individual basis as well.

\begin{table}[htbp]
  \centering
  \caption{Average accuracy and inference times for a varying number of activities to be identified}
  \label{tab:avg_accuracies_inference_times}
  \resizebox{0.9\columnwidth}{!}{
  \begin{tabular}{ccc}
    \toprule
    \textbf{User Activities} & \textbf{Average Accuracy} & \textbf{Average Inference Time (s)} \\
    \midrule
    2 & 94\% & 0.69 \\
    3 & 92\% & 0.81 \\
    4 & 89\% & 1.06 \\
    5 & 94\% & 1.54 \\
    6 & 92\% & 3.16 \\
    \bottomrule
  \end{tabular}
  }
\end{table}

\subsubsection{\textbf{RQ5}} \textbf{``How much'' VR traffic data do we need to capture in order to accurately identify VR applications and user activities?}

\textbf{Insight 5.} \textit{The model maintains over 91\% accuracy for identifying VR applications and 88\% for specific user activities even with less than 10 minutes of collected traffic for each application.}

In Figure~\ref{fig:actacc}, we present results on accuracy, precision, and recall while training our random forest model with varying percentages of the dataset we have collected (training our model with 20\%, 40\%, 60\% and 80\% of the collected data as well as the entire dataset we have collected). The data was collected across 25 distinct VR apps, spanning a total of 15 hours, 49 minutes, and 38 seconds. The goal of this extensive data collection procedure was to ensure that a broad range of user interactions within these VR environments was represented in a dataset that was comprehensive enough for analysis. Our results show that even with 20\% of the collected dataset (i.e., less than 10 minutes of collected traffic for each application), we can identify VR applications with an accuracy of more than 91\% and specific VR user activities with an accuracy of more than 88\%. Furthermore, we notice that increasing the size of the training dataset increases accuracy, precision, and recall but not substantially (less than 3\%.). In general, this indicates that even with smaller training datasets, the classifier had ``enough'' data in order to learn patterns of the network traffic that helped accurately classify applications and activities.



\begin{table}[htbp]
  \centering
 
    \caption{System performance metrics for random forest and decision tree classifiers}
    \label{tab:dtsysmet}
    \resizebox{0.8\linewidth}{!}{
      \begin{tabular}{lcc}
        \toprule
        \textbf{Metric} & \textbf{Random Forest} & \textbf{Decision Tree} \\
        \midrule
        Total Training Time (s) & 96.71 & 77.65 \\
        CPU Usage (\%) & 62.27\% & 11.72\% \\
        Memory Usage (GB) & 1.38 & 1.37 \\
        \bottomrule
      \end{tabular}
    }
 \end{table}

\subsubsection{\textbf{RQ6}}\textbf{What are the performance and systems trade-offs of using machine learning for VR application and user activity identification?} 

\textbf{Insight 6.} \textit{As we increase the dataset size, the model accuracy correspondingly improves, with a little to no change in average computation and memory overhead.}

To answer this RQ, we quantified systems related metrics during model training and inference, such as CPU, GPU, and memory usage, as well as training and inference times. We made use of Google Colab's computing resources, which included a combination of CPU and GPU resources. The CPU of the system we used was a single physical Intel(R) Xeon(R) CPU @ 2.20GHz, with two logical and one physical CPU. The GPU of the system we used was an NVIDIA Tesla T4 with 16GB of memory. Google Colab gave us the option to train the random forest and decision tree classifiers on CPU resources and the deep learning model on GPU resources.

In Table~\ref{tab:avg_accuracies_inference_times}, we present results on the identification accuracy and inference time when trying to identify a varying number of activities (ranging from two to six activities) for a given VR application. Our results indicate that regardless of the number of activities to be identified for a given VR application, the identification accuracy is at least 89\%. At the same time, as we increase the number of activities to be identified, the inference time increases. Overall, the inference time ranges between 0.69-3.16 seconds.

 \begin{table}[htbp]
  \centering
    \caption{Deep learning model performance metrics}
    \label{tab:deep_learning_metrics}
    \resizebox{0.75\linewidth}{!}{
      \begin{tabular}{lcc}
        \toprule
        \textbf{Metric} & \textbf{Activity} & \textbf{Application} \\
        \midrule
        Total Training Time (s) & 142.70 & 143.27 \\
        Average GPU Utilization (\%) & 13\% & 14\% \\
        Power Draw (W) & 29.06 & 27.48 \\
        Memory Used (MB) & 165.96 & 166.06 \\
        \bottomrule
      \end{tabular}
    }
 
\end{table}

In Tables~\ref{tab:dtsysmet} and~\ref{tab:deep_learning_metrics}, we present results on the system performance for all the models we used. The training time for the random forest and decision tree classifiers was 96.71 seconds and 77.65 seconds respectively when the entire collected dataset was used. The usage of CPU resources was rather light: on average, 28.75\% and 11.72\% of the available CPU resources were occupied while training the random forest and the decision tree classifiers on the entire collected dataset. The memory used during training was also not significant (10.9\% and 10.81\% of the available memory resources for random forest and decision tree respectively). In the case of the deep learning model, the total training time was 142.70 seconds and 143.27 seconds when training for activity and application identification respectively. The average GPU utilization was 13\% and 14\% while training for activity and application identification respectively. The power was 29.06W and 27.48W in the case of training for activity and application identification respectively, while the memory used was 165.96MB and 166.06MB respectively.

In Table~\ref{tab:performance_metrics}, we present systems performance metrics for the random forest classifier while varying the size of the training dataset to demonstrate how the training dataset size impacts those metrics. Our results show that the training time increases with the size of the training dataset, while the CPU and memory usage do not considerably change. Specifically, when trained with 20\% of the collected dataset, the training time was about 12 seconds, increasing to about 96 seconds when the entire dataset is used for training.

\begin{table}[htbp]
    \centering
    \caption{Random forest classifier performance metrics for different percentages of the collected dataset used for training purposes.}
    \resizebox{1\columnwidth}{!}{
    \begin{tabular}{cccc}
        \toprule
        \textbf{Dataset Size} & \textbf{Training Time (s)} & \textbf{CPU Usage (\%)} & \textbf{Memory Usage (\%)} \\
        \midrule
        20\% & 25.57 & 65.04 & 12.33 \\
        40\% & 44.02 & 65.95 & 12.28 \\
        60\% & 64.17 & 67.06 & 12.53 \\
        80\% & 79.14 & 61.17 & 12.60 \\
        100\% & 96.71 & 60.67 & 12.74 \\
        \bottomrule
    \end{tabular}
    }
    
    \label{tab:performance_metrics}
\end{table}

\section{Related Work}

In this section, we discuss prior related work. Prior work has focused on privacy and security aspects of VR, VR performance related issues, the use of ML for classification of network traffic, VR user experience issues, and the use of ML for quality of service issues in VR. Our work is different from prior work, since we explore whether network traffic generated by VR headsets/applications can be used to breach the privacy of VR users by identifying the VR applications that are used as well as the activities performed by VR users in the context of these applications. 

\subsection{Security and Privacy in VR}

In the realm of security and privacy, there are papers that contribute to the understanding of the privacy risks and data collection practices within VR environments. OVRseen~\cite{trimananda2022ovrseen} has presented a comprehensive study on privacy policies within the Oculus VR platform, aiming to raise awareness of privacy risks in VR applications. In addition, a framework to identify users based on sensor data, highlighting the importance of protecting user privacy in VR environments has also been proposed~\cite{jarin2023behavr}. To evaluate security, privacy, and safety threats in VR learning applications, a risk assessment framework has been proposed, which makes use of attack trees and highlights the significance of addressing these threats in order to guarantee users a safe and private educational experience \cite{8651847}.

Another study explores the complex moral conundrums raised by VR technologies, highlighting the significance of taking social, physiological, and ethical factors into account when designing a system \cite{8558774}. To ensure the responsible and sustainable development of VR applications across various industries while addressing the ethical implications and safety concerns associated with VR technologies, this study emphasizes the need for extensive research on the psychological effects of VR, particularly on children.
Researchers have further discussed the ethical challenges arising from the convergence of VR and social networks, highlighting the threats to privacy and autonomy and proposing an Aristotelian approach to address these concerns through policy recommendations for policymakers, providers, and users \cite{OBrolchain2016-dr}.

\subsection{VR Related Performance Analysis}

Prior work has also explored the performance analysis of VR applications. Coterie \cite{meng2020coterie} is a framework designed to enhance multiplayer VR experiences by reducing network bandwidth requirements. There are also studies that delve into the analysis of network protocols, infrastructure, system performance, and resource usage in VR, identifying technical challenges and proposing solutions to improve scalability, latency, and overall user experience~\cite{cheng2022we}~\cite{trimananda2022ovrseen}. Particular research has highlighted the difficulties and methods for gathering, processing, and presenting dynamic performance data \cite{471180}. They also cover the design and implementation of Avatar, a data-immersive virtual environment intended for performance analysis and real-time adaptive control of application behavior in high-performance computing systems. The design, implementation, and analysis of a multi-user VR application are presented by Parthasarathy \textit{et al.} along with assessing network performance and user experience based on real and simulated users, as well as the effect of tick rates on throughput and round trip times \cite{9148390}. A non-intrusive methodology for evaluating commodity VR systems has been also explored~\cite{10.1145/2964284.2967303}. This methodology offers insights into how 3D scene complexity affects timing accuracy and how positioning accuracy metrics trade off sensitivity and precision.

\subsection{Network Traffic Classification} 

Techniques and tools have been presented in the field of network traffic classification to effectively categorize network traffic, particularly when data capture and ML are integrated with Wireshark \cite{8319360}, which offers robust features for capturing real-time network data. ML techniques have been applied to online applications utilizing real-time network traffic classifiers \cite{a14080250, alshammari2021}.
Online classifiers for network traffic that use Naive Bayes, k-Nearest Neighbors, and decision trees, have shown a substantial degree of accuracy in identifying network traffic \cite{a14080250}. The use of ML techniques to identify malicious network traffic (anomalies) has also been explored by Alshammari \textit{et al.} \cite{alshammari2021}. This work offers a method for successfully identifying and categorizing various kinds of network threats by combining long short-term memory and a convolutional neural network.

\subsection{User Experience} 

Previous studies have examined components of the VR application user experience, including comfort, presence, immersion, and usefulness. Metrics like presence, workload, usability, and flow were used in the assessment process to analyze user experience in VR simulations of driving and walking scenarios to determine how different VR devices affect user performance and satisfaction~\cite{RHIU2020103002}. User participation, affective reactions, and cognitive strain  has also been examined in fully immersive VR settings \cite{8613667}. They emphasize the value of interaction, user interface design simplicity, and the potential for VR technology to improve user experience. For the purpose of comparing environmental ratings and user responses, a multi-method approach has been investigated that proposes the use of VR as an empirical research instrument to evaluate user experiences in a real building and an associated virtual model \cite{KULIGA2015363}. 

\subsection{Use of ML for Quality-of-Service in VR} 

Prior work has investigated ML methods for analyzing VR-related data, especially the quality-of-service provided to VR users. Particularly, a framework has been proposed that combines ML, echo state networks, and multi-attribute utility theory to optimize resource allocation for wireless VR users while taking tracking accuracy, transmission latency over small cell networks, and processing delay into account~\cite{8254650}. The quality of experience in the case of 360-degree VR movies and the effects of different parameters on it have also been explored and ML models have been used to predict and optimize the quality of experience~\cite{9163348}. In addition to providing a thorough model for assessing VR quality of service with an emphasis on tracking accuracy, transmission delay, and processing delay, a framework has been proposed that integrates VR applications into wireless cellular networks along with a learning-based resource management algorithm~\cite{8395443}.

\section{Conclusion}

In this paper, we investigated the impact that network traffic generated by VR headsets and applications can have on the privacy of VR users. Specifically, we made use of ML models to learn unique patterns of the network traffic generated by VR applications, which allowed us to accurately identify VR applications and the activities conducted by VR users in the context of these applications. It is consistent with research indicating that random forest models perform better than decision trees, which supports the idea that ensemble methods offer superior generalization. Nonetheless, this outcome emphasizes the significance of capturing feature interactions in the VR context—something that random forests are better at than single decision trees. The performance of the model was largely influenced by network throughput metrics, especially those that recorded brief spikes in data transmission that are representative of activities unique to virtual reality. We summarize our findings below: 

\begin {itemize}

\item Specific VR applications and VR user activities can be identified with high accuracy through off-the-shelf ML models.

\item Assuming that attackers can create a dataset that contains a set of VR user activities, these activities can subsequently be identified in network traffic traces captured from ``unknown'' VR applications. 

\item VR applications and VR user activities can be accurately identified even when attackers have access to relatively small training datasets containing only a few hours of VR network traffic in total (less than 10 minutes of network traffic for each VR application).

\end{itemize}

Our findings highlight that major concerns about the privacy of VR users can arise when attackers simply collect the (even encrypted) network traffic that VR headsets generate and make use of off-the-shelf ML models to learn characteristics of the generated network traffic.  

\bibliographystyle{ACM-Reference-Format}
\bibliography{sample-base}

\end{document}